# The irregular (integer) tetrahedron as a warehouse of biological information

Tidjani Négadi

Physics Department, Faculty of Science,
University of Oran, 31100, Oran, Algeria;
E-mail: tnegadi@gmail.com

**Abstract:** *A "variable geometry" classification model of the 20 L-amino acids and the 20 D-amino acids, based on twenty, physically and mathematically, labeled positions on tetrahedrons, and extending Filatov's recent model, is presented. We also establish several physical and mathematical identities (or constraints), very useful in applications. The passage from a tetrahedron with (possibly) maximum symmetry to a tetrahedron with no symmetry at all, here a distinguished integer heronian tetrahedron, which could "describe" some kind of symmetry breaking process, reveals a lot of meaningful biological numerical information. Before symmetry breaking, and as a first supporting result, we discover that the L- and D-tetrahedrons together encode the nucleon-content in the 61 amino acids of the genetic code table and the atom-content in the 64 DNA-codons. After a (geometric) symmetry breaking, and also an accompanying (physical) "quantitative symmetry" restoration concerning atom numbers, more results appear, as for example the atom-content in this time 64 RNA-codons (61 amino acids and three stops), the remarkable Downes-Richardson-shCherbak nucleon-number balance and, most importantly, the structure of the famous protonated serine octamer $Ser_8+H^+$ (L- and D- versions), thought by many people to be a "key player" in the origin of homochirality in living organisms because of its unique property to form exceptionally stable clusters and also its strong preference for homochirality. Using all the labeling possibilities, we find the more fundamental neutral serine octamer $Ser_8$ (L- and D-versions). We also revisit, in this paper, the number 23! which is at the basis of our recent arithmetic approach to the structure of the genetic code. New consequences, not yet published, and also new results, specially in connection with the serine octamer, are* *given. Finally, a remark on the inclusion the "non-standard" versions of the genetic code, in the present formalism, is made.*

# 1. Introduction

This paper is devoted, first, to a new classification of the twenty amino acids based on the *heronian (integer) tetrahedron*, including and extending the recent one by Filatov (Filatov, 2009) based on the usual tetrahedron and includes also the twenty mirror-image D-amino acids. We start with the regular, *achiral*, tetrahedron, the most basic of all polyhedra (the first of the five platonic solids) with maximum symmetry and end up with the *irregular*, *chiral*, *integer*-heronian tetrahedron, noted symbolically "**117**", with no symmetry at all. In this way we have, besides a "symmetry-breaking" process "begeting chirality", a coherent chiral-framework classification of the amino acids, distributed in the faces, vertices and edges of the chiral *integer* heronian tetrahedron (and also its mirror-image for the 20 D-amino acids). Recall that in the case of the tetrahedron the framework could not always be chiral and the amino acids which are disposed on it are chiral (except of course glycine). The *whole object that is the chiral heronian tetrahedron-and-chiral amino acids on it is, in this way, chiral*. At the same time, a slight "quantitative symmetry-breaking", at the level of atom number and inherent to the Filatov's ordinary tetrahedron (extended) model, with only *physical* labeling (nucleon, carbon, hydrogen and atom numbers), is cured and the latter restored. Another nice virtue of the above "**117**" heronian tetrahedron is that it incorporates (encodes) naturally and strikingly, throught its geometric characteristics the correct numeric structure of the famous protonated serine octamer $Ser_8+H^+$, thought to be a promising actor in the origin of homochirality (Cooks et al., 2001; Hodyss et al., 2001). It is precisely the passage from a regular tetrahedron, with possible maximum symmetry, to the heronian (irregular) tetrahedron, with no symmetry at all (no two edges equal), that is a symmetry breaking, that leads to the serine octamer. By introducing a supplementary and concomitant *mathematical* labeling, for the 20 positions (amino acids) in the vertices, edges and faces of the tetrahedron, the so-called "generalized Plato's Lambda" numbers or *Tetraktys*, its more fundamental neutral form $Ser_8$ is revealed. This is examined in section 2. Our second aim, in this paper, and in section 3, is to revisit the number 23!

which was at the basis of our arithmetic model of the genetic code (Négadi, 2007, 2008, 2009) but in the context of the present work. We show, in particular, that it incorporates, too, the above serine octamer structure and encode also several other mathematical characteristics of the genetic code. As this paper is intended first for the readers of Symmetry: Culture and Science, and maybe other people at large, we include two Appendices to (i) define the physical (chemical) and mathematical entities used, (ii) ease reading and (iii) render possible skill/verifications by the reader. In the first, we collect detailed chemical data concerning the 20 amino acids as well as the DNA (RNA) units Thymine (Uracil), Cytosine, Adenine and Guanine. In the second, we give some mathematical definitions concerning certain (elementary) arithmetic functions used in this paper (and in others), that help to unveil the hidden "biological information". In particular, famous Euler's totient function φ is briefly presented, with here interesting applications.

## 2. An extended classification model of the amino acids and the serine octamer $Ser_8$

From about seven hundred known amino acids only exactly twenty of them are coded by the genetic code. It is also a recognized fact that life uses *exclusively* the L-form amino acids to make proteins and *exclusively* the D-form sugars in the backbones of DNA and RNA. However, and this is a guenine paradigm shift (or turning point) in biology, the D-amino acids are also used in many living beings, from bacteria to mammals (see for example Yang et al. 2003), in various biological strategies. In fact, L or D is a matter of convention, the two forms (called also enantiomers) exist and some people say that life could even have begun "blind" to the D- or L-forms of the amino acids, maybe used both, and latter "chosed" one of them, the L-form, by inventing "control quality", proofreading mechanisms and "checkpoints". In the contemporary ribosome, it is mainly the Aminoacyl-t-RNA synthetases that play a central role for the correct handedness (L-amino acids) because the other component, the Peptidyl Transferase Center and the codon-anticodon Decoding Center, are "blind" to the handedness of the amino acids. The Decoding Center guarantees only the identity of an amino acids but not its handedness. Enantiomers have very similar physical properties, identical with respect to ordinary chemical reactions, and the difference arise only when they interact. In the recent years, the field of D-amino acids chemistry and biochemistry has grown and these have now the same status as their L-forms. In the following, therefore, we shall consider an extended classification comprising the 20 L-amino acids as well as the 20 D-amino acids.

We start from Filatov's classification model of the twenty amino acids, which is a consequence of his symmetric *table of the genetic code*, and based on the *tetrahedron*. The author does not precise the geometric nature of the tetrahedron (regular or irregular) but any tetrahedron has 4 *faces*, 4 *vertices* and 6 *edges*. At one side, we have the *regular* tetrahedron with 24 isometries and, at the other we have the *irregular* tetrahedron with 7 possible isometries. The extreme case where *all* the edges of the tetrahedron are *different* corresponds to a complete loss of symmetry and we are precisely interested in this case, in this work. The 20 amino acids are disposed on the tetrahedron as shown in the figure 1 below (see Filatov, 2009). Four amino acids A, N, L and F are on the center of the four faces (blue), four amino acids G, P, K and Y (green) are on the four vertices and the remaining twelve amino acids (in red) are disposed on the 6 edges (2 amino acids per edge). Filatov has discovered a "*quantitative*" symmetry at the level of the *nucleon number* (the Hasegawa-Miyata Parameter (HMP), Hasegawa, and Miyata, 1980). The nucleon (proton or neutron) is the basic building-block of (ordinary) matter.

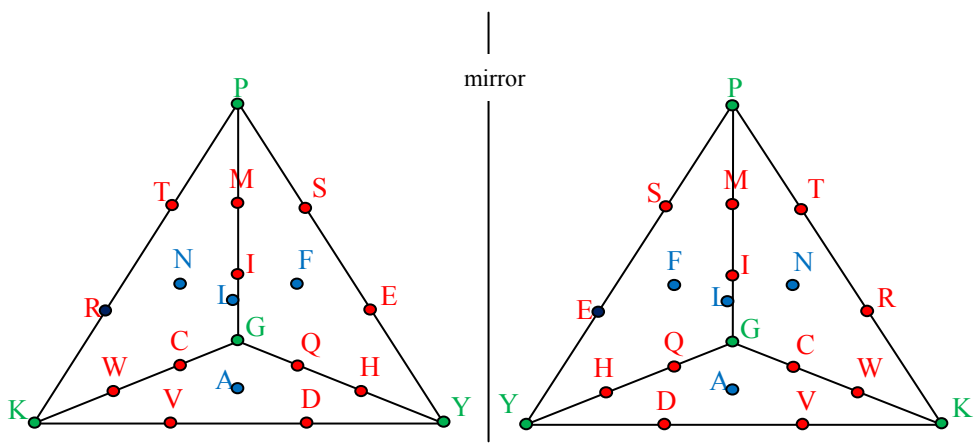

**Figure 1**: The L-amino acids (ex. left) and the D-amino acids (right); the vertical line: the mirror.

The sum of the nucleon numbers of the amino acids (side-chains only) in the two pairs of faces is (for one tetrahedron, the L-tetrahedron say):

$$628+627=626+629=1255 \qquad (n)$$

The nucleon numbers of the 20 amino acids are given in Appendix 1. Here, the two pairs of faces are defined in the figure by (PKY/YKG) and ((PGY/PGK), respectively. We have used Filatov's tetrahedron model to deduce the existence of the equivalents of equation (n) for carbon and for hydrogen. For atom number, there is, a-priori, no (exact) quantitative symmetry, as for carbon and hydrogen, but we shall see below that this "quantitative symmetry-breaking" could be cured, thanks to the heronian tetrahedron mathematical properties, and the (quantitative) symmetry restored, with interesting consequences. For carbon, it writes (for the respective faces, as in (n),

$$33+34=32+35=67 \qquad (c)$$

Using an elementary (funny) manipulation of numbers, by adding the "digits" of the numbers in Eq.(n) in base-100 gives 34+33=32+35, which is Eq.(c) up to a trivial permutation. The nucleon and carbon atom numbers appear therefore, somehow, "linked" as we have already experienced earlier (see Négadi, 2009) for the total number of nucleons in the 20 amino acids (12+55=67). Now, for hydrogen, we have

$$64+53=62+55=117 \qquad (h)$$

That each pair of faces gives the (correct) nucleon, carbon and hydrogen numbers for the 20 amino acids is not at all trivial. As a matter of fact, in each pair of faces, there is *only* 16 different amino acids, with four of them situated on an edge, contributing *two times*, and these 16 amino acids give the correct number of nucleons, carbon and hydrogen numbers for the 20 (different) amino acids. Thus, there is something, not trivial, at work. If now we consider atom numbers, we have for the two pairs 108+97=205 and 104+99=203, which is not 204, the number of atoms in the 20 amino acids and the "quantitative" symmetry is *broken*. However, their mean gives the correct value: (205+203)/2=204. We shall see below how this "quantitative symmetry-breaking" for atom numbers could be "restored".

There exist also a *secondary* "quantitative" symmetry at the level of *carbon* (C), *hydrogen* (H), *atom* and *nucleon* numbers but now between the four vertices (G, P, Y, K) and the four face centers (L, A, F, N):

$$\text{\#C-atoms}^{(v)} = \text{\#C-atoms}^{(f.c.)} = 14 \tag{1}$$

$$\text{\#H-atoms}^{(v)} = \text{\#H-atoms}^{(f.c.)} = 23 \tag{2}$$

$$\text{\#atoms}^{(v)} = \text{\#atoms}^{(f.c.)} = 39 \tag{3}$$

$$\text{\#nucleons}^{(v)} = \text{\#nucleons}^{(f.c.)} = 221 \tag{4}$$

We could also take Eqs.(1)-(4), collectively, and represent this "secondary" quantitative symmetry by the sum of the four numbers in the following interesting identity:

$$297 = 297 \tag{5}$$

The *identities* (or constraints) such as (n), (c) and (h) will be called in the following *primary* and those in (1)-(5) *secondary*. We shall show below the importance of the relation in Eq.(5) concerning serine and for other interesting consequences. As mentioned in the introduction, the above tetrahedron(s) could also be labeled but the numbers of the 3-D generalization of famous Plato's Lamda, or *3-D Tetraktys*[1] The numbers on the four faces, when these latter are taken separately, are given as shown below

```
      1                    1                   1                    8
     2  4                 3  4                2  3                12  16
    4  8  16             9  12  16           4  6  9             18  24  32
   8 16 32 64           27 36 48 64         8 12 18 27           27 36 48 64
```

Let us note, first, that the sums of the numbers in each of the above four number-patterns are as follows: 155, 220; 90, 285, respectively. Now, when the tetrahedron is reconstructed, that is when the four "faces" are assembled, some numbers become redundant so that the sum of all theses numbers is only 350 (see Phillips, ref. in footnote 1, and figure 2 below). In this way, we could establish, also for the Tetraktys, the equivalent of Filatov's nucleon number "quantitative" symmetry (see Eq.(n), same order):

---

[1] Stephen M. Phillips Plato's Lambda-its meaning, generalization and connection to the Tree of Life (Article 11, http://smphillips.8m.com).

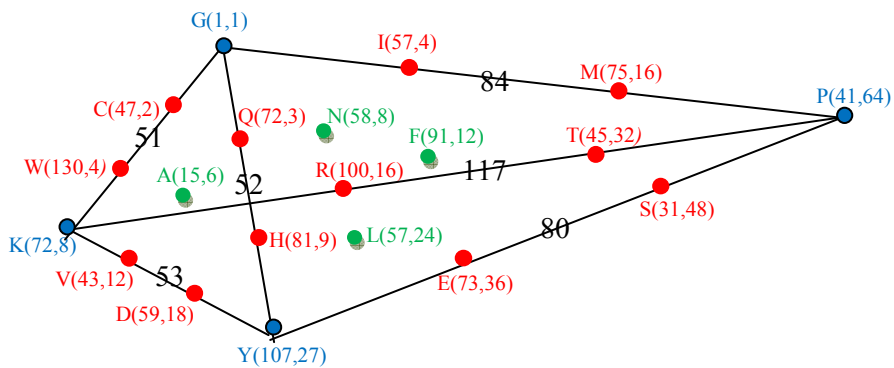

**Figure 2**: The heronian tetrahedron "117" hosting the 20 amino acids

$$285+90=220+155=375 \tag{T}$$

This Tetraktys-number identity could be shown to be also derivable from the above secondary identities relations (1)-(5). As a matter of fact, by adding (5) to the sum of the prime factors of (1) through (4), or their $a_o$-functions (see Appendix 2), we have (*two times*)

$$297+a_o(14)+a_o(23)+a_o(39)+a_o(221)=375 \tag{6}$$

Let us now return to the heronian tetrahedron which was mentioned several times above. This interesting mathematical object (shown in figure 2 above), also called "perfect pyramid" (Buchholz, 1992), has *integer* edges, faces areas and volume. We have represented in the figure only one heronian tetrahedron, say the L-heronian tetrahedron, but its mirror-image D-heronian tetrahedron is also considered, but here not represented. Also, for each amino acid, the first number in the parenthesis is the number of nucleons and the second the Tetraktys number. Now from a large family of heronian tetrahedrons, known in mathematics, the integer heronian tehtrahedron having the *smallest maximum* edge-length is the one with edge-lengths 51, 51, 53, 80, 84, 117. As mentioned in the introduction, this tetrahedron is often noted "**117**", in reference to its special property (smallest maximum edge-length). As mentioned in the introduction, the above (irregular) tetrahedron has all its edges *different* so that it is *chiral*. The faces corresponding to the order in equation (n) have as (integer) areas 1800, 1170, 1890, 2016. The volume V is equal to 18144 (see Buchholz, 1992). As a first application, consider the integer volume. Mathematically the volume of the mirror image of V is –V, via the Cayley-Menger determinant but this would have no impact because the arithmetic functions we use here are insensible to the sign of an integer, as they are only the sum of the prime factors. We have $a_0(V)=a_0(-V)=29$ so that $a_0(a_0(V)+a_0(-V))=31$, SOD(V)=18 and $B_0(V)=56$, where SOD is the Sum Of Digits and the functions $a_0$ and $B_0$ are defined in Appendix 2. This is *serine*; it has 31 nucleons in its side-chain, 56 in the block of its "residue" and 18 in the water molecule to complete the block. (Note that 56+18=74 is the number of nucleons in a block, the same for 19 amino acids and, as it is well known, proline is an exception.) Serine is also "present" when considering the Tetraktys numbers with sum 350. Looking at the numbers in the detail, we see that the numbers 4, 8, 12 and 16, and only these, appear two times or with multiplicity 2. This gives for the total sum 80+270=350 where 80 is for the above four numbers, having multiplicity 2, and 270 corresponds to the rest of the numbers, with no multiplicity. We have $B_0(270)=31$ and $B_0(270)+B_0(80)=56$ so that $2 \times B_0(270)+B_0(80)=31+56=87$ is equal to the nucleon number in the "residue" of serine, 105-18, where 18 is for the nucleons in a water molecule as mentioned above. This latter could be found several ways and two of them are the following $B_0(375-350)=18$ or, simply $a_0(375)=18$, where 375 is the sum of the Tetraktys numbers in the two pairs of faces (see Eq.(T)). These three numbers 31, 56 and 18, characterizing serine, are identical to the above three ones, obtained from the heronian tetrahedron.

The heronian tetrahedron has as the sum of its six integer edges 437 but when the *two pairs* of faces are considered each edge contributes twice so that the total sum is 2×437=874. This number identifies nicely with the number of nucleons in the protonated serine octamer $Ser_8+H^+$, in a state called "maximum exchange" where it "catches" exactly 33 hydrogen nuclei or nucleons. As a matter of fact we could write for the total number of nucleons for eight serines, the octamer, and its "charging" proton, on the one hand, and the 33 "exchangeable" hydrogens, on the other, 8×(31+74)+1+33=841+33=874. (Each serine has 4 exchangeable hydrogens **H**$_2$N- (amine group), **H**O- (hydroxy group) and –COO**H** (acid group) and a charging-carrying proton, that is 4×8+1=33, hydrogen atoms.) [2] In experiments serine clusters can appear between the mass ratio m/z=841 (no exchange) and m/z=874 (complete exchange). It is interesting, and remarkable, that the sum of all the edges of the two pairs of faces of the above heronian (L-)tetrahedron, 874, could be made to reproduce numerically the details of the protonated octamer and its 33 exchangeable hydrogens. It suffices to introduce the *primary* identity (c), for carbon, in the form 33+34-(32+35)=33-33=0, that selects the number 33: 874-33+33=841+33=874. Selecting instead the number 34 (33+charging proton), we obtain the neutral serine octamer $Ser_8^0$ and the rest in protons (hydrogen atoms) 840+1+33=874. Also, selecting from the six edges the two equal maximum lengths, we get 2×117+640=874 and, introducing this time the *secondary identity* for carbon, in Eq.(1) and assuming that the first primay identity is already applied, we end up with 248+593+33=74. The first two numbers are respectively the number of nucleons in the 8 side-chains of the 8 serines forming the octamer (8×31) and the number of nucleons in their 8 blocks as well as the charging proton (8×74+1=592+1). The serine tetramer (4 serines) is the smallest serine cluster known to exhibit homochiral preference (Cooks et al., 2001). It is also thought that the octamer could be made of two tetramers (Nemes, 2005, Costa and Cooks, 2001). As only a speculative exercise, and following the reasoning about the serine octamer, the tetramer ($Ser_4+H^+$,) would have 4×4=16 "exchangeable hydrogens" and, adding a "charging" proton, we would end with a tetramer in "maximum exchange" having a number of nucleons equal to 4×105+1+16=421+16=437. It is remarkable that this number half 874 or the sum of the six edges of the heronian tetrahedron (51+52+53+80+84+117). The introduction of the identity (5) in 437 gives 140+297=437 which has a simple meaning: the first term, 140, is the number of nucleons in the four side-chains and the sixteen "exchangeable hydrogens" 4×31+16 and the second, 297, is the number of nucleons in the four blocks and the "charging" proton 4×74+1. This is in agreement with the fact that the octamer is thought to be made of two tetramers. Finally, the D-octamer is reproduced the same way, using this time the mirror image, or D-tetrahedron.

Now, we come back to the "quantitative symmetry-breaking", mentioned above. We had for the two pairs 108+97=205 and 104+99=203. In fact, the discrepancy comes from the NOS-atoms (nitrogen, oxygen and sulfur), as for hydrogen and carbon the symmetry is obeyed. In the detail, and for the four faces, we have 108=64+44, 97=53+44, 104=62+42 and 99=55+44, where in each case the first number refers to hydrogen number and the second to the CNOS-atoms. There exist numerical patterns that suggest that hydrogen is in some way separated from the other four atoms (see the next section) and we shall keep this separation working in the above four relations where the numbers of CNOS-atoms are, respectively 44, 44, 42 and 44. These latter correspond to the faces "628", "627", "626" and "629", respectively, and we shall show below that their atom numbers are rather given by the $B_o$-functions of the areas of theses faces. The three edges of these faces are (53, 80, 117), (51, 52, 53), (51, 84, 117) and (52, 80, 84), respectively. The (integer) areas of these faces have, as mentioned above, the values 1800, 1170, 1890 and 2016. The formula for the area (Heron of Alexandria, 10-75 A.D.) is simply the square root of s(s-a)(s-b)(s-c) where s=(a+b+c)/2 is the *semiperimeter*. Now, the $B_o$-functions of the areas are calculated to be $B_o(1800)=42$, $B_o(1170)=45$, $B_o(1890)=43$ and $B_o(2016)=44$, respectively. Adding the respective hydrogen atom numbers gives 64+42=106, 53+45=98, 62+43=105 and 55+44=99, respectively. Finally, grouping the two pairs of faces apart gives

$$106+98=105+99=204. \tag{a}$$

---

[2] I would like to thank Lars Konermann (Department of Chemistry, The University of Western Ontario, Canada) for very kind explanations concerning the exchangeable hydrogens.

Thus, by replacing the CNOS-number of each face of the tetrahedron by the $B_o$-function of the area of that face, the "quantitative symmetry" at the level of atom-number is *restored*, as it was mentioned before (remember, the quantitative symmetry for hydrogen is not broken 64+53=62+55=117).
Now, we come to some interesting applications. Consider, first both (ordinary) tetrahedrons, the L-tetrahedron and the D-tetrahedron (see the figure). The 20 positions on each one of them are, first, labeled by the *nucleon numbers* of the amino acids. We know that there is a "balance" between two pairs of faces (primary identity (n)) and four "balances" between the four vertices and the four centers of the faces (secondary identities (1)-(4) or (5)) involving the carbon, hydrogen, atom and nucleon numbers. Finally, each tetrahedron involves 4 *vertices*, 4 *centers* and 6 *edges* or 14 "invariant" objects. Writing down the sum for the two tetrahedrons (L- and D-) gives

$$2\times(2\times1255+2\times297+14)=2\times2\times(1255+297+7) \tag{7}$$

The first expression, at the left of the equal sign, means 2 tetrahedra (L- and D-) and the second, at the right, means twice two pairs of faces and, in each pair, 7 (concomitantly) means 2 faces sharing 5 edges or 2+5=7. Writing, first, 1255 as 627+628, 297 as 221+76 and 7 as 3+4 (or 2+5) and, second, noting that the different parts are independent, we could for example group in the parenthesis the *odd* numbers together and the *even* ones together, to obtain the following interesting partition

$$2\times2\times(627+221+3)=3404 \tag{8}$$

$$2\times2\times(628+76+4)=2832 \tag{9}$$

The first equation gives the total number of *nucleons* in the 61 amino acids and the second the total number of *atoms* in 64 DNA-codons. As a matter of fact, there are 145 nucleons in the 5 quartets, 188 in the 3 sextets, 660 in the 9 doublets, 57 in the triplet and 75+130=205 in the two singlets (see Appendix 1) and therefore 145×4+188×6+660×2+57×3+205×1=3404 nucleons, in 61 amino acids. For 64 DNA-codons, each one of the four nucleobases T, C, A and G appears 48 times and there are therefore 48×(15+13+15+16)=2832 atoms (see Appendix 1). Now, taking instead 1255=626+629 in (8) and (9), the result would be 3404+8=3412, for the first, and 2824, for the second, so that it is essentially the same result as above. Note also that 3412 is the number of nucleons in 61 amino acids where these latter are in their "physiological" state, that is when some few amino acids are charged (Downes and Richardson (2002), shCherbak, 2008). Concerning the number 3412, mentioned above, Downes and Richardson (2002) have shown the existence of an exact *nucleon-number balance* between the 61 amino acids *residue molecules* (in their "physiological" state) side-chains and the 61 blocks (57×56+4×55) where the unique amino acid proline has 55 nucleons instead of 56 which is the number of nucleons in the remaining 19 amino acids. (Note that in this case the number of nucleons in the 20 amino acids is 1256 or 1255+1.) To see the balance, we consider the three identities (n), (c) and (a), where in this latter we add in both members 90+90=90+90 (corresponding to the blocks). We have

$$\text{L-tetrahedron} \rightarrow 2\times(1255+67+384)=3412 \tag{10}$$

$$\tag{11}$$

   D-tetrahedron → 2×(1255+67+384)=3412

and the equality of the two tetrahedra could "fit" (or encodes) the remarkable *Downes-Richardson nucleon number balance*. Downes and Richardson (Downes and Richardson, 2002, see also shCherbak, 2008) considered that some few amino acids (in their "physiological" state) are charged and also the case where proline has 42 nucleons in its side chain and 73 nucleons in its block and showed that there are 3412 nucleons in the 61 side-chains and 3412 (=57×74+4×73) nucleons in the corresponding 61 blocks. We have also, recently, established this very numerical balance, by using other mathematical technics (Négadi, 2011). Now, we have already established the restoration of the "quantitative" symmetry of atom number. Consider therefore the sum Σ=2×(1255+67+117+204) including the nucleon, carbon, hydrogen and atom numbers in the two pairs of faces of the tetrahedron. We have (Σ=2×31×53)

   $B_0(\Sigma)$=86+31=117                (12)

Where 86 corresponds to the sum of the prime factors, $a_0(\Sigma)$, and the rest, 31, to the sum of the prime indices and the big Omega function (or the number of prime factors, here 3; see Appendix 2). The result is the total number of *hydrogen* atoms in the 20 amino acids in agreement with the pattern "**16**+**7**=**23**" (see below): 86 hydrogen atoms in 5 quartets, 9 doublets and 2 triplets (21+50+7+8) and 31 hydrogen atoms in 3 sextets and 1 triplet (22+9). Now, we include the blocks, in the number of atoms and collect everything for *one* pair of faces: Eqs.(n), (5), (c), (h) and (a) to which we add 2×90=180, for the blocks, and also the number of edges (5) and centers of faces(2), that is 7. We have for the sum of the prime factors and their indices for the sum, 2127 (=3×709; 709 is the 127th prime), of all the mentioned quantities

   $A_0$(1255+297+67+117+384+7)=841        (13)

This number, again, corresponds to the protonated serine octamer. The selection of the sub-set comprising the four last numbers above is also interesting. Computing the quantity

   2×2×$A_0$(67+117+384+7)=4×48=192        (14)

for both tetrahedrons, L- and D-, that is for 2×2 pairs of faces, we find this time the total number of nucleobases in 64 codons (each of the 4 nucleobase appears 48 times). Next, we introduce the quantity $Q_o=2\times1255+2\times375+(1170+1800+1890+2016)$ comprising the nucleon numbers, the Tetraktys numbers and also the (integer) areas, the two pairs of faces of the heronian tetrahedron, the L-tetrahedron, say. We have

$$B_o(Q_o)=248 \tag{15}$$

The nucleon numbers and the Tetraktys numbers for the four vertices and the four face centers gives $\Phi^{(v, f.c.)}=2\times221+150=592$. Adding the two quantities above gives

$$B_o(Q_o)+ \Phi^{(v, f.c.)}=248+592=840 \tag{16}$$

The relation (16) corresponds exactly to the *neutral serine octamer* $Ser_8^o$: there are 248 nucleons in its eight side-chains (8×31) and 592 nucleons in its eight blocks (8×74). $B_o(Q_o)$ and $\Phi^{(v, f.c.)}$ give therefore these two parts. Also, taking the $A_o$-function of these two numbers, we obtain

$$A_o(248)+A_o(592)=112 \tag{17}$$

Amazingly it appears that 112 is equal to the number of *atoms* in the neutral serine octamer, as serine has 14 atoms (side-chain and block) and 8×14=112. Finally, considering the two tetrahedra (L- and D-), we have $B_o(2\times112)=32=31+1$ and this corresponds to the protonated monomer $Ser^+$ (side-chain only). In the equations (16) and (17) nothing prevents from adding the number of tetrahedron(s) involved, here 1, to get 840+1=841 (nucleons) and 112+1=113 (atoms) for the protonated serine octamer. For the two tetrahedra we could form the product $[B_o(112)+1]\times[B_o(112)+1]$ and this latter is equal to $29^2=841$, again to protonated serine octamer. Now, we form the two following quantities

$Q_1 = 2 \times 1255 + (1170 + 1800 + 1890 + 2016)$ (18)

$Q_2 = 2 \times 375 + (1170 + 1800 + 1890 + 2016)$ (19)

where, in $Q_1$, the nucleon and the heronian tetrahedron numbers are considered and, in $Q_2$ the Tetraktys numbers replace those of the nucleons. First we have $A_o(Q_1) + A_o(Q_2) = 76 + 104 = 180$ which coud be written as 114+66. This is the carbon-content in the 61 amino acids: 114 in 5 quartets, 9 doublets and 2 singlets ($4 \times 9 + 2 \times 33 + 3 + 9$), on the one hand, and 66 in the 3 sextets and the triplet ($6 \times 9 + 3 \times 4$), on the other. This is also in agreement with the pattern "**16**+7=**23**" (see above and below). Now, we use the $B_o$, instead. We have in this case $B_o(Q_1) + B_o(Q_2) = 80 + 108 = 188$. This is the number of nucleons in the three sextets serine, leucine and arginine. As $Q_1 = 2 \times 13 \times 19^2$ and $Q_2 = 2 \times 3 \times \mathbf{31} \times 41$, we get immediately **31**+157 which selects serine's number of nucleons, 31. Some little additional manipulation gives the final result **31**+57+100, respectively. Taking the total sum, $180 + 188 = 2^4 \times 23$, we have $a_o(368) = 31$. The sum of the prime indices and the number of factors give 18, the water molecule. Equivalently $B_o(368) = 31 + 18 = 49$. Finally, crossing the four values, we introduce the numbers 156=76+80 and 212=104+108 from which we have $A_o(156) + A_o(212) = 105$. This is the nucleon-number of serine (side-chain and block). As we have already shown, here, in this paper, these *same* numbers refer to serine; in particular the difference $A_o(156) + A_o(212) - B_o(368)$ gives the right number of nucleons in the block of the residue of serine (31+74=49+56=87+18=105).

To end this section, let us return to the two important, and interesting, numbers 248 and 840. Both correspond to the serine octamer, in the first the block is not included. We show below that, in fact, the number 248 begets the number 840 and shares with the heronian tetrahedron rare mathematical properties. As a matter of fact, *the number 248 is the smallest number >1 for which the three Pythagorean means, the arithmetic, geometric and harmonic means of Euler's totient-function $\varphi$ and the divisor-function $\sigma$ are all integers*. We have $\varphi(248) = 120$ and $\sigma(248) = 480$. As for the means we have the a-mean=(120+480)/2=300, the g-mean=$\sqrt{120 \times 480} = 240$ (integer!, see the equation below) and finally the h-mean=($2 \times 120 \times 480$)/(120+480)=192. From all these relations we have

$\varphi(248) + \sigma(248) + \text{g-mean}[\varphi(248), \sigma(248)] =$ (20)

120+480+240=248+592=840

In the last step, we isolated the proper divisors of 248 and wrote σ(248)=248+232. This result is the same as in Eq.(16), the neutral serine octamer. We have thus seen that the complete serine octamer is derivable from the side-chain part and the number 248 seems therefore more fundamental than 840. It is even "hidden" in the L/D-classification based on the regular tetrahedron. In Eq.(7), the right-hand side is equal to $2^2 \times 1559$ which is the factorization of the sum 6236 (3404+2832). It appears that the sum of the prime-indices is equal to 248 (1559 is the 246$^{th}$ prime). Now, $A_0(840)=33$ and find that 840, itself, begets the 33 "exchangeable" hydrogens (see above). The sum of the three numbers in the chain 248→840→33 (for one tetrahedron) leads to $B_0(1121)=105$ and, considering the two tetrahedra, L- and D-, we have $A_0(2 \times 1121)=106=105+1$. These are respectively the neutral monomer of serine and the protonated monomer, experimentally observed. The number 192 which is the total number of nucleobases in 61 codons appears, here, two times: (i) $\varphi(840)=192$ and (ii) g-mean[$\varphi(248)$, $\sigma(248)$]=192, see above). This could describe 61 RNA-codons and 61 DNA-codons, inasmuch as 840 could also reveal the four DNA units {T(126), C(111), A(135), G(151)} and the four RNA units {U(112), C(111), A( 135), G(151)} where, in parenthesis, the number of nucleons in each nucleobase is given. As a matter of fact, using Eq.(16), we get by adding 840 and its φ-function (see section 3)

$$840+\varphi(840)=1032 \tag{21}$$

This last number is identical with the total nucleon-sum of the eight units: 523 for the DNA units and 509 for the RNA units. The number 1032 could be written 442+590, by isolationg the nucleon identity (4), that is the *physical term* $2 \times 221$, and next by applying the carbon identity (c), 67-67=0, we find (442+67)+(590-67)=509+523 which is the result. Considering also the two tetrahedra and using the results following Eqs.(18)-(19), we have

$$2 \times 1255 + 2 \times (B_0(Q_1)+B_0(Q_2))=2 \times (1255+188)=2 \times 1443=2886 \tag{22}$$

This is also an interesting result, as it gives the total number of atoms in 61 RNA-codons, 2560, and in 3 stops, 326 (see Rakočević, 2009). Rakočević used the *nucleotides* for the 61 RNA-codons and the *ribonucleosides* for the 3 stop-codons (UAA, UAG and UGA): 45U+48C+44A+46G giving **2560** atoms and 3UMP+4AMP+2GMP giving 326 atoms. Let us note that 326 could be written as 128 for the nucleotide-part, or the "side-chain", and 198 for the ribose/phosphate-part, or the "block". Let us also write one copy of 1255 as 245+1010, according to the pattern "16+7" (see above and below) where 245=188+57 (3 sextets S, L, R, and the triplet I). Knowing that 188 is also equal to $B_o(Q_1)+B_o(Q_2)$=108+80, we could therefore rewrite (22) as (108+57)+(2×188+80+1010+1255)=2886. It is now sufficient to apply the primary identity (c) for carbon to get (108+**57**+33)+(2721-33)=198+2688. We have therefore found the "block"-part, 198, see above. Using now, in 2721, the decomposition of 188 as **57**+131 and using the secondary identity for carbon (1), we have (198+128)+2560=326+2560, that is the correct partition into codons and stops (see above and Rakočević, 2009). Is is quite astonishing that we could also reveal the structure of serine using nucleon- and atom-numbers in all the chemical engredients amino acids, DNA and RNA. Take for example the two numbers 2886 and 2560. Their difference, 326, was associated above to the 3 stops and they are both relevant: $A_o(2886)$=67+9=76, the number of carbon atoms in 23 Amino Acids Signals and $a_o(2560)$=18+5=23, precisely these 23 AASs, (). Also, their difference (the three stops) gives $A_o(326)$=204 which is the number of atoms in the 20 amino acids (see also Rakočević, 2009). Consider therefore the quantity $Q_3$=3404+2886+2560=2×3×5$^2$×59. We have $B_o(Q_3)$=74+31=105, precisely the detailed structure of serine (see above). Add now to $Q_3$ the number of atoms in 64 DNA codons $Q_4$=$Q_3$+2832 (see Eq.(9)) to get $A_o(Q_4)$=105. Also we have that $B_o(½Q_4)$=106, the number of nucleons in the protonated monomer of serine, 105+1. Finally $Q_5$=3404+2832+2560=$2^2$×3×733 (733: 130$^{th}$ prime) leads to $A_o(Q_5)$=874 which corresponds to the protonated serine octamer in its state of maximum exchange (see above in this section). Finally, the two main numbers of the tetrahedron 437 and 874 are also informative, by themselves. First, $B_o(437)$ written as $A_o(437)+\Omega(437)$=59+2=61 has a nice interpretation with respect to the genetic code mathematical structure: 59 codons for degenerate amino acids and one codon for each one of the two singlets (M and W). Interestingly, the sum of the $a_o$-functions of the four semiperimeters 125, 78, 126 and 108 of the heronian tetrahedron gives also 61. As for 874, closely related to the serine octamer, and homochirality, as we have seen in this paper, we have $B_o(874)$=65 and this last number is conspicuously identical with the number of what are known as the "*biological*" *space groups* which survive, out of a total of 230 crystal groups when considering *chiral* molecules (see for example Mainzer, 1996 ).

## 3. Revisiting 23!

In 2007, (Négadi, 2007), we have designed an arithmetic model of the genetic code based on the number 23!. From its twofold, decimal and prime-factorization representations in equations (II) and (III)

$$23! = 1 \times 2 \times 3 \times 4 \times \ldots \times 21 \times 22 \times 23 \tag{I}$$
$$23! = 25852016738884976640000 \tag{II}$$
$$23! = 2^{19} \times 3^9 \times 5^4 \times 7^3 \times 11^2 \times 13 \times 17 \times 19 \times 23 \tag{III}$$
$$(I)-(II)=0, \tag{IV}$$
$$(I)-(III)=0 \tag{V}$$

we have deduced the multiplet structure of the (standard) genetic code and computed the right degeneracies as well as many other interesting results (Négadi, 2008, 2009). Every digit, from 1 to 9 in Eq.(II), was associated to an amino acid coded by more than one codon. Two zeros were associated to the two singlets methionine and tryptophane and three zeros to the three stop codons. One of them, for example, gives the total number of hydrogen atoms in the 61 amino acids as $a_0(23!)+\Omega(23!)+117=200+41+117=358$, where 117 is the sum of all the digits in (II) as well their number (zero excluded). This last number is nothing but the number of hydrogen atoms in the 20 amino acids and, the remaining part, 200+41=241, corresponds to the 41 degenerate codons (amino acids). It appears that we did not fully exploit all the numerical facets hidden in 23!. For example, an unnoticed nice result comes from the above two representations, (II) and (III), and consists in adding *all* the (47) digits, as "individual", even ignoring the place value (for example count 11 as 1+1) and also the factorial-sign. In this way we obtain $99+74+2\times(2+3)=183$, which identifies nicely with the total number of nucleobases in the 61 codons inasmuch as the prime factorization of 183, $3\times61$, is "taylor-made": 61 codons and 3stops, because $a_0(183)=61+3=64$, and also 18 amino acids with degeneracy and 2 non-degenerate singlets, because the sum of the prime-indices SPI(183)=18+2=20. This is not the end, because if we consider the first "representation", (I), which is in fact the definition of the factorial, and proceed as above for its (individual) digits, we get 114 which identifies with the number of nucleobases in 38 codons (3 nucleobases per codon). Substracting 114 from 183 gives 69 the number of nucleobases in 23 codons. We obtain therefore the pattern 23+38 for the 61 codons. Note that the sum 183+114 gives 297. (114: 73 from 1 till 16 and 41 from 17 till 23).

Another interesting result comes from the three representations of 23! written above as (I), (II) and (III). In the table below, we compute the sum of all the digits, and their number, needed to write each representation (excluding zeros and ignoring place-values, exponent positions, etc., just the digits)

|     | #digits (zero excluded) | sum |
| --- | --- | --- |
| I   | 35 | 114 (**73**+41) |
| II  | 18 (**11**+7) | 99 |
| III | 20 | 74 |

The total sum is 360 and we have, adding its number of divisors $360+\tau(360)=360+24=384$. This number is equal to the number of atoms in the 20 amino acids (side-chain and block). In the table the number 114 is also written according to the partition 73+41, already mentioned. Moreover, the number of digits in the decimal place-value representation is also partitioned according to the parity of the digits (even/odd): 18=11+7. Using these two partitions, we have

(i) (**73**+**11**)+300=84+300=384

and

(ii) (99+74+**7**)+(**73**+41+**11**+20+35)+$\sigma(360)$=180+204=384.

The case (ii) corresponds to the blocks and the side-chains, respectively, and case (i) describes the partition into the set comprising the 5 quartets, the 9 doublets and the 2 singlets, 16 amino acids having 76+177+20+27=300 atoms, on the one hand, and the 3 sextets and the triplet, 7 amino acids having 62+22=84 atoms, on the other. Let us denote this pattern by "**16**+7=**23**"; we shall meet it again later in the next section (see Appendix 1). The number of atoms, 384, could be derived otherwise. As a matter of fact, including the sum of the three 23s in the first members of Eqs.(I)-(III) and the nine "missing" zeros, including those of Eqs.(IV)-(V) which are "necessary" to express the equivalence of the three relations (I), (II) and (III), we get 360+(5+5+5)+9=384. Pushing the computation to its extreme, by including the number of digits in the three 23s, 6, we end up with 390, a very useful number, see below. Above, the total number of hydrogen atoms in the 61 amino acids was computed as $a_o(23!)+\Omega(23!)+117=200+41+117=358$, where 117 is the sum of all the digits in (II) as well their number (zero excluded). Now, if we include the five zeros, we have 99+23=122 and, adding the sum of all the digits in (I), 114, we get 236. This is the number of CNOS-atoms (carbon, nitrogen, oxygen and sulfur) in the 61 amino acids. In this way we find the total number of atoms in the 61 amino acids 236+358=594, 236 for CNOS and 358 for hydrogen H. Returning to the above number 390, it appears that it is just equal to the number of atoms in the 41 *degenerate* codons (amino acids) and, substracting it from 594, gives 594-390=204, which is the number of atoms in 20 amino acids. Now, we turn to another type of consequences from the number 23!, that were not published before. We introduced (Négadi, 2009) the use of certain (useful and generating information) simple algorithms known as mathematical "black-holes" to "produce" biological information. One of them, call it the "black-hole$^{(123)}$"-algorithm, where the "black-hole" is the number 123, works as follows: (i) start with *any* number and count the number of even digits and the number of odd digits, (ii) write them down next to each other (by concatenation) following by their sum. Treat the result as a new number and continue the process. This latter is very quick even for big numbers, as 23!. The first iteration, using (II), with 16 even numbers and 7 odd numbers gives 16723 (16+7=23) and this first iteration agrees with the pattern "**16**+**7**", mentioned several times in the first section. The second iteration gives 235, the third 123 and finally the fourth (check) also 123.

$$23! \to \mathbf{16723} \to \text{csod}(16723)=\mathbf{1}$$
$$\downarrow \qquad\qquad\qquad \downarrow$$
$$\text{SPI}(16723)=359 \to 235 \to 123 \to 123 \to \mathbf{840}+1=841 \to 841+A_o(840)=874$$
$$\downarrow$$
$$a_o(a_o(16723))=603 \to 235 \to 123 \to 123 \to \mathbf{1443}\ (+1)$$
$$\varphi^{(3)}(16723)=1440 \to \mathbf{1443}$$

As the number 16723 is a five-digits number, but with special consideration (see above), we shall treat it differently from the rest of iterations which are all three-digits numbers. It is capable to give, as its consequences, two numbers (i) 359 as the sum of the prime-indices of its prime-factors and (ii) 603 as the sum of the prime factors of it's $a_o$-function (see the schematic diagram above). We see that the sum of the four numbers in the second line give 840, the neutral serine octamer number. Adding the complete sum of the digits, 840+csod(16723), gives the protonated form 840+1=841. Finally, adding the $A_o$-function of 840 leads ot 840+csod(16723)+$A_o$(840)=841+33=874, the serine octamer in the state of "maximum exchange, see above. Now, we use the φ-function and consider exactly some iterations applied to 16723. We get, at the third iteration $φ^{(3)}(16723)=1440$. If we just add to this last number the number of iterations, here, 3 we have 1440+3=1443. This number is not unknown; it is the number of nucleons in the 23 AASs 1255+188 (see above). The addition of csod(16723)=1, or 1443+1, could even fit the usual case where proline has 41+1 nucleons. The sum $a_o(a_o(16723))$+SPI(16723)+[$φ^{(3)}(16723)$+3]+235+2×123=2×1443=2886 gives the same result as in Eq.(22), that is the total number of atoms in 61 RNA-codons (2560) and in 3 stops (326), at the sole condition to introduce the carbon identity, used several times above, into SPI(16723) and write it 359-33+33=326+33. Rearranging the terms we end up with 2560+326=2886. We have shown in the previous section how the numeric structure of the serine octamer arises from the mathematical characteristics of the heronian tetrahedron. It could shown that the Tetraktys, too, is able to reveal this strange structure. First, we have to take the two tetrahedrons in order to include L- as well as D-amino acids. The number 350 is the sum of all the numbers situated at the 20 places, as explained above, and 375 corresponds to the (same) sum on the two pairs of faces (see above). These four numbers are 155, 220, 90 and 285. The following relation

$$(2×350+A_o(155)+A_o(220)+A_o(90)+A_o(285))+B_o(375)$$
$$=(700+140+1)+33 \quad (23)$$
$$=(840+1)+33$$
$$=874$$

which is constructed from the Tetraktys numbers only, gives again the same result for the serine octamer as in the first section, from the heronian tetrahedron, or as above from the "black-hole[123]" algorithm. We speculate that there could maybe exist a "link" between these different approaches. For example we have that the number of *partitions* of the number 23 is equal to 1255 (procedure numbpart(.) in softwares). It appears also that the number of partitions of the number 20 is equal to 627. The difference leads to 1255-627=628 or 627+628=1255 and this relation is precisely one of the two filatov's identities mentioned at the beginning of section 2 (see Eq.(n)). Another interesting result comes when we take the ratio between the number of divisors of 23! and the φ-function of the number of partitions of 23:

$$\frac{\tau(23!)}{\varphi(\text{numbpart}(23))}=192 \tag{24}$$

The result, 192, is equal to the total number of nucleotides in 64 codons and, as φ(840) is also equal to 192, we get 2×192=384, a nice starting point for establishing the multiplet structure of the genetic code, another way, that is *from one and small number* (Négadi, 2011). As a last word, let us remark that a close (and strange) relation exists between serine, which was in the focus, in this paper, and proline, also known to be a quite singular amino acid (ex. "entangled" side-chain and block). Notably, it has also been shown that chirality has significant impact on the assembly of proline clusters (Myung et al.**,** 2006). Looking at that acute apex of the heronian tetrahedron in Figure 2, where proline resides, we have that the sum of the nucleon number of proline's side-chain, 41, and its associated Tetraktys number, 64, is equal to 105=41+64, which is precisely the number of nucleons in serine (side-chain and block). Also, by calling, again, the carbon identity (c), we get (41+33)+(64-33)=74+31 which is serine, in the detail, 31 for the side-chain and 74 for the block. Inversely, and first, the positive difference between the numbers for serine, 48 and 31 is equal to 17 which fits the atom-number of proline. Second, take now the total sum of the nucleon numbers on the tetrahedron, 1255, and the Tetraktys numbers, 350, that is 1605 (3×5×107), and compute the sum of the prime factors of this sum, we find 3+5+107=8+107=115 which is precisely the total number of nucleon in proline. Introducing the carbon identity (33-33=0) gives 41+74; this is the correct partition into side-chain and block. Moreover, adding the sum of the prime-indices of the prime factors and their number (Ω-function) to the numbers for serine, 31 and 48, gives 31+48+36=115. Finally, writing 36 as 8+28, we have (48+8+31)+28 and by introducing the atom number (secondary) identity (1), we get 73+42=115. This is the other form of proline's nucleon number: 41+1 in the side-chain and 74-1 in its block. The "manifestation" of serine and its clusters seems not to be something linked to the tetrahedron classification of the 20 amino acids considered in this paper, alone. We have studied recently the small set of the amino acid *precursors* and found a similar "manifestation" of serine, its clusters, in particular the octamer (Négadi, 2011).

We end this paper by considering the possibility to include the "nonstandard" genetic codes, mentioned at the end of the introduction. We have seen in this paper, that a prominent role is given to the number of amino acids 61 and also to the 3 stops, *i.e.*, the *standard* genetic code. A raised question by a reviewer was whether something could be said, in the present model formalism, about the experimentally known "nonstandard" genetic codes which are however also known to concern only very few living organisms, compared to the standard, or quasi-universal genetic code. This is an interesting question and we shall show that indeed something could be said. As a matter of fact, we start from the number of nucleotides in the 61 coding (or sense) codons 183=3×61, *i.e.*, three nucleotides per codon. This number (derived by us several times, as the one in section 3 also mentioned in Appendix 2) has as the prime factorization 3×61 so that its $a_0$-function (sum of the prime factors) is 61+3=64 with, here, an immediate new "interpretation": 61 amino acids sense codons and 3 stop-codons. Please note in this latter case *the "new" function of the number 3, as the number of "stops" for the "standard" genetic code* which we recall concerns the great majority of the living organisms. Now, some 18 "nonstandard" genetic codes have been discovered these last years (see Elzanowski and Ostell, 2010 for a recent updated compilation). Looking at these genetic codes tables, we have that the *sense* codon number oscillates between 60 and 63 with only four possible (observed) cases 60, 61, 62 and 63. Equivalently, the possible number of stop-codons oscillates between 1 to 4, with also four possible cases 4, 3, 2 and 1. In general, in these "re-assignments" and without intering into the details of the biochemical machinery, a sense codon could become a stop codon and, conversely, a stop codon could become a sense codon, coding for some "new" amino acid (in the same canonical set of 20 amino acids or even comprising the 21th or the 22th amino acids Selenocysteine or Pyrrolysine). It is precisely at this point that, one more time, Euler's totient function $\varphi$, and also $\sigma$, the sum of divisors function, could help to "describe" these features. Take, as the starting point, the numbers 61 and 3 for the "standard" genetic code case where the former refers to the sense codons and the latter to the stops. They are both prime (see above) and we have

- $\varphi(61)=60=61-1$
- $\sigma(61)=62=61+1$
- $\varphi(3)=2=3-1$
- $\sigma(3)=4=3+1$

For a prime p, recall that φ(p)=p-1 and σ(p)=p+1. It is not difficult to see that by introducing the above relations, in the "standard form" 63+3=64, three and only three other emerging cases, (ii)-(iv), are possible. In Summary, we have

- (i) 64=61+3
- (ii) 64=60+4
- (iii) 64=62+2
- (iv) 64=63+1

These four relations seemingly describe *all* the following 18 observed cases (and also the case of Selenocysteine and Pyrroloysine too) where the number of stop-codons is indicated in the parenthesis (see Elzanowski and Ostell, 2010): The Standard Code (3), The Vertebrate Mitochondrial Code (4), The Yeast Mitochondrial Code (2), The Mold, Protozoan, and Coelenterate Mitochondrial Code and the Mycoplasma/Spiroplasma Code (2), The Invertebrate Mitochondrial Code (2), The Ciliate, Dasycladacean and Hexamita Nuclear Code (1), The Echinoderm and Flatworm Mitochondrial Code (2), The Euplotid Nuclear Code (2), The Bacterial, Archaeal and Plant Plastid Code (3), The Alternative Yeast Nuclear Code (3), The Ascidian Mitochondrial Code (2), The Alternative Flatworm Mitochondrial Code (1), Blepharisma Nuclear Code (2), Chlorophycean Mitochondrial Code (2), Trematode Mitochondrial Code (2), Scenedesmus Obliquus Mitochondrial Code (3), Thraustochytrium Mitochondrial Code ( 4) and Pterobranchia Mitochondrial Code (2). It appears that the inclusion of all these genetic code variants could be extended also at the level of the *number of nucleotides*, itself, beyond the number of codons, thanks again to the φ-function. We shall develop these new results in a forthcoming paper.

**Appendix 1**

In this appendix we give some numeric data concerning the 20 amino acids and the DNA and RNA units, used in the text. In the Table below, the detailed atomic composition of the amino acids *side-chains* is given: H for hydrogen, C for carbon c, N for nitrogen, O for oxygen and S for sulfur. Also, the atom and nucleon numbers atre given. The 20 amino acids are organized into the five known multiplets of the *standard* genetic code: 5 quartets (M=4), 3 sextets (M=6), 9 doublets (M=2), 1 triplet (M=3) and 2 singlets (M=1); the multiplicity M gives the number of codons. *The amino acids are given in the one-letter usual code* (in parenthesis) *and only the numbers for the side chains are given.* When one considers also the blocks, then the corresponding number for the (common) block must be added; for example for the number of atoms one must add 9, for the nucleons 74, etc.. Second, the number of *atoms* in the five nucleobases (or nucleotides) is as follows Uracil (U, 12)/Thymine (T, 15), Cytosine (C, 13), Adenine (A, 15) and Guanine (G, 16), see the picture below (courtesy from Dr. Gary E. Kaiser). When the "block", made of the ribose sugar and phosphate, is added to the nucleotides the *ribonucleosides* have the following content in atoms UMP($C_9H_{13}N_2O_9P$, 34), CMP($C_9H_{14}N_3O_8P$, 35), AMP($C_{10}H_{14}N_5O_7P$, 37), GMP($C_{10}H_{14}N_5O_8P$, 38), (see Rakočević, 1997).

| M | Amino acid | H | C | N/O/S | #Atom | #Nucleon |
|---|---|---|---|---|---|---|
| 4 | Proline (P) | 5 | 3 | 0 | 8 | 41 |
|   | Alanine (A) | 3 | 1 | 0 | 4 | 15 |
|   | Threonine (T) | 5 | 2 | 0/1/0 | 8 | 45 |
|   | Valine (V) | 7 | 3 | 0 | 10 | 43 |
|   | Glycine (G) | 1 | 0 | 0 | 1 | 1 |
| 6 | Serine (S) | 3 | 1 | 0/1/0 | 5 | 31 |
|   | Leucine (L) | 9 | 4 | 0 | 13 | 57 |
|   | Arginine (R) | 10 | 4 | 3/0/0 | 17 | 100 |
| 2 | Phenylalanine (F) | 7 | 7 | 0 | 14 | 91 |
|   | Tyrosine (Y) | 7 | 7 | 0/1/0 | 15 | 107 |
|   | Cysteine (C) | 3 | 1 | 0/0/1 | 5 | 47 |
|   | Histidine (H) | 5 | 4 | 2/0/0 | 11 | 81 |
|   | Glutamine (Q) | 6 | 3 | 1/1/0 | 11 | 72 |
|   | Asparagine (N) | 4 | 2 | 1/1/0 | 8 | 58 |
|   | Lysine (K) | 10 | 4 | 1/0/0 | 15 | 72 |
|   | Aspartic acid (D) | 3 | 2 | 0/2/0 | 7 | 59 |
|   | Glutamic Acid (E) | 5 | 3 | 0/2/0 | 10 | 73 |
| 3 | Isoleucine (I) | 9 | 4 | 0 | 13 | 57 |
| 1 | Methionine (M) | 7 | 3 | 0/0/1 | 11 | 75 |
|   | Tryptophane (W) | 8 | 9 | 1/0/0 | 18 | 130 |
|   | Total | 117 | 67 | 9/9/2=20 | 204 | 1255 |

**The 20 amino acids atomic composition**

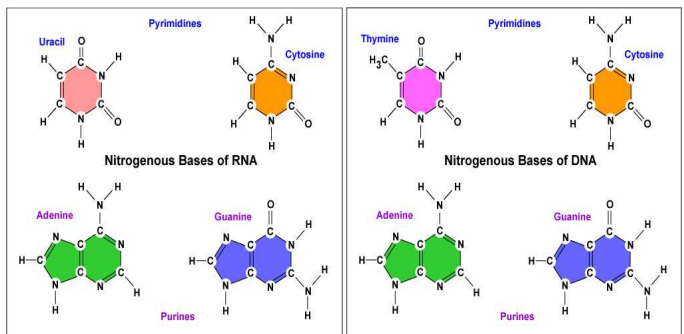

**The nitrogenous nucleobases of RNA and DNA** (From Gary E. Kaiser's Microbiology Home Page, with his permission.)

**Appendix 2**

In this appendix, we give the definition of some *elementary* mathematical tools, used in this paper. First, we use the Fundamental Theorem of Arithmetic which states that every natural number n could be written, uniquely, as a product of primes each raised to a given exponent $n=p_1^{a_1} \times p_2^{a_2} \times p_3^{a_3} \times p_4^{a_4}\ldots$. For a given number n the arithmetic function $a_o(n)$ gives the *sum of the prime factors of n*, *including* multiplicity. When the multiplicities are discarded, the corresponding function is called $a_1(n)$. We also define the function, $A_o(n)$ to be the sum of $a_o(n)$ *and* the sum of the *prime-indices* of the prime factors. The big-Omega function $\Omega(n)$ counts the *number of the prime factors*. We also define the function $B_o(n)$ as $A_o(n)+\Omega(n)$. Let us give an example. Take the number n=183, mentioned in section 3. Its prime decomposition is 3×61, the prime-indices of the two prime factors 3 and 61 are respectively 2 and 18 and $\Omega(183)=2$. We have $a_o(183)=3+61=64$, $A_o(183)=a_o(183)+2+18=84$ and $B_o(183)=A_o(183)+\Omega(183)=86$. We also use famous Euler's totient or φ-function (a fundamental mathematical object in Cryptography) which gives the total number of numbers that are *co-prime* to n and, specially for a prime p, it is simply p-1). In other words, it gives a count of how many numbers in the set {1, 2, 3, …, n} share no common factors with n that are greater to one. For any number n, the general formula $\varphi(n) = p_1^{a_1-1}(p_1-1)\ p_2^{a_2-1}(p_2-1)\ldots p_n^{a_k-1}(p_k-1)$ could be used to compute φ, by hand, but one could also use more quickly computer software as phi(n) and sigma(n) in Maple6, used here.